\documentstyle[times,pramana,epsf,floats]{ias}
\begin{document}
\mark{{Dictionary of LHC Signatures}{A. Belyaev et al.}}
\title{Dictionary of LHC Signatures\footnote{contribution to the WHEPPX
workshop, Chenai, January 2008}}

\author{A. Belyaev$^1$, I.A. Christidi$^2$, 
     A. De Roeck$^3$, R.M. Godbole$^4$, 
     B. Mellado$^5$, A. Nyf\-fe\-ler$^6$,
     C. Petridou$^2$, D.P. Roy$^7$}
\address{
$^1$School of Physics \& Astronomy, University of Southampton,
Southampton SO17 1BJ, UK;
Particle Physics Department, Rutherford Appleton Laboratory,
Didcot, Oxon OX11 0QX, UK.
\\
$^2$ 
Division of Nuclear and Particle Physics, Physics Department, Aristotle
University of Thessaloniki, Thessaloniki, 54 124, Greece.
\\
$^3$ CERN, 1211 Geneva 23, Switzerland and University of Antwerp, Belgium.
\\
$^4$ 
Centre for High Energy Physics, Indian Institute of Science,
Bangalore, 560012, India.
\\
$^5$ 
Physics Department,
University of Wisconsin - Madison,
Madison, Wisconsin 53706 USA.
\\
$^6$ 
Regional Centre for Accelerator-based Particle Physics,
Harish-Chandra Research Institute,
Chhatnag Road, Jhusi,
Allahabad - 211 019, India.
\\
$^7$ 
Homi Bhabha Centre for Science Education, Tata Institue of Fundamental Research,
Mumbai-400088, India.
} 

\keywords{LHC, Dark Matter, Discrimination , Underlying Theory}

\abstract{
We report on a plan to establish a ``Dictionary of LHC Signatures",
an initiative that started at the WHEPPX workshop in January 2008.
This study aims towards the strategy on distinguishing of 3 classes of dark
matter motivated scenarios such as R-parity conserved
supersymmetry, Little Higgs models with T-parity conservation and
Universal Extra Dimensions with
KK-parity for  generic cases of their realization
in wide range of the model space. Discriminating signatures are
tabulated and will need a further detailed analysis.

}
\maketitle

\def\met{{\slash\!\!\!\!\!\:E}_T}

\def\GeV{\ifmmode {\mathrm{\ Ge\kern -0.1em V}}\else
                   \textrm{Ge\kern -0.1em V}\fi}%
\let\gev=\GeV

\def\TeV{\ifmmode {\mathrm{\ Te\kern -0.1em V}}\else
                   \textrm{Te\kern -0.1em V}\fi}%
\let\tev=\TeV

\section{Introduction}
The particle physics community is eagerly awaiting the start-up of 
 the LHC. The measurements at this proton-proton collider with a center
of mass system energy of 14 TeV will 
  shed light on  the origin of 
electroweak symmetry breaking and are expected to provide
collider signatures of dark matter (DM), thus directly revealing 
new physics beyond the Standard Model (BSM). 
 
The identification of
BSM signals at the LHC and establishing the underlying theory will become 
a central question, after the discovery. Correctly   
identifying the new physics scenario from the data will be a very important
task, and due to the  very many 
possible  scenarios it is likely to be  a very difficult or 
perhaps even unsolvable puzzle.
However, among the many   compelling BSM scenarios proposed so far,
only a few  provide a stable  DM candidate (with a correct relic density) 
and at the same time solve the hierarchy 
and fine-tuning problem of the SM Higgs sector.
Hence we turn our attention in this paper to those BSM models that 
fulfill these requirements. 

The idea of this study, which started off at the Workshop on High Energy Physics
Phenomenology (WHEPP X) in January 2008, is to design a
strategy on how to distinguish three representative BSM candidates, namely
supersymmetry with conserved R-parity (SUSY)~\cite{stdsusyref}, the Littlest 
Higgs model~\cite{LH_original} with T-parity (LHT)~\cite{T_parity}  
and Universal Extra 
Dimensions with KK-parity (UED)~\cite{Appelquist:2000nn}. In fact, for
all these models, one expects very similar event topologies at the LHC,
with new particles produced in pairs which then subsequently decay in (long)
cascades to the lightest stable DM particle which escapes detection.
For each scenario we choose generic regions 
in the parameter space, each characterized by specific features of
the DM particle properties. The regions selected are 
allowed 
by the cosmological constraints on the relic density; e.g.\ for SUSY
this means the so-called bulk, 
co-annihilation, focus point 
and resonant annihilation region (funnel corridor).

The final goal of this study is to classify
generic properties and signatures of each class of models
and find the strategy for discriminating the underlying model.
In this paper we report the plan towards this final goal and present
qualitative arguments for the different signatures that will be used.
This classification and the strategy are discussed in the next section.

Similar questions have been 
studied in the context of the so called inverse 
problem of supersymmetry at the LHC~\cite{ArkaniHamed:2005px}, 
and footprints for SUSY models~\cite{Kane:2007pp}. 
A recent study~\cite{Hubisz:2008gg}
aims to discriminate SUSY, and to a lesser extent also LHT and UED models,
using a variety of different kinematical observables
related to the spin difference of the underlying theories, by using tailored
benchmark points particularly suitable for  the LHC start-up.
In the present study we extend the classes of various observables
and will attempt to establish a strategy 
for more generic regions of the parameter space for every
class of these BSM scenarios.
The results  themselves will be reported in a follow up report, 
following the full study
which will also take into account experimental issues,
applied to the comprehensive list of observables listed below.

\section{Generic LHC signatures of the BSM and their powers of discrimination} 
Generic properties and signatures of the
SUSY, LHT and UED models are the following.

\noindent 
I) \underline{Spin statistics}:
SUSY superpartners have a spin different  compared to their partners,
while LHT and UED are theories with "bosonic" supersymmetry, {where the
 SM particle and its heavy partner have the same spin. This difference
can be probed effectively by the following observables:}
\begin{itemize} 
\item
\underline
{Difference of the total cross section}:
This has been discussed, e.g., in Refs.~\cite{Choi:2006mr,Datta:2007xy,Kane:2008kw}.
It was shown in ~\cite{Datta:2007xy}
that the cross section of chargino-neutralino production
in SUSY is typically one order of magnitude lower than the cross section
of the analogous particle production ($W_H Z_H$) in LHT. Note however,
that for total cross-sections one needs to  control the theoretical 
uncertainties, such as parton distributions, renormalization and 
factorization scale uncertainties etc. Alternatively, one needs
to find effects which may be less sensitive  to these uncertainties.
The experimental issues of relevance to this
measurement are the systematics in the luminosity measurement, the lepton 
identification and trigger efficiency, the jet energy scale and energy and 
momentum resolution. Note that the experimental cuts can modify the expected
relative rates of the different models.
\item
\underline
{Various angular correlations between final 
state  particles}: This issue has been discussed, e.g. in 
Refs.~\cite{Hubisz:2008gg,Barr:2004ze,Barr:2005dz,Datta:2005zs,Athanasiou:2006ef,Hooper:2006xe}.
The invariant lepton mass distributions as well as the
lepton-quark invariant mass
distributions  were shown to be capable of  discriminating between 
SUSY and UED models, even for similar masses of the heavy partners 
in both  the models~ \cite{Datta:2005zs}.
{Since a direct spin measurement is impossible due to the LSP in the final 
state, such correlation studies are the only handle. However, this is a
very challenging measurement.  
Choice of a particular final state as well as that of particles 
therein  to study the correlations plays a crucial role, 
particularly since 
the combinatorics can sometimes completely smear out the 
differences. The angular, energy and momentum resolution of the measurement 
 also 
plays a very important role.}
\item
\underline
{Polarization of the final state SM particles}: Polarization
of the top quarks and taus,  is reflected in their decay products and 
is does experimentally
accessable. {The same experimental issues that  affect the
study of angular correlations are important here as well.} 
The polarization may  be used to determine the character of the 
DM particle and hence the underlying model 
parameters~\cite{Nojiri:1994it,Guchait,Calibbi:2007uw} 
as well as to  sharpen the search strategies~\cite{Guchait:2002xh}. In 
stau-coannihilation region of SUSY, the final state 
signatures will be exhibited by  very soft $\tau$ leptons. 
In this case, the polarization of the $\tau$ can be used very effectively to 
reduce the background from QCD jets~\cite{ggr}.
\item 
\underline
{Difference in the  direct and indirect DM detection rates:}
{The DM detection rates in the  DM search experiments 
can play a very important role in this discrimination between models as 
discussed, e.g., in 
Ref.~\cite{Hooper:2006xe,Altunkaynak:2008ry,Arrenberg:2008wy}}.
{The ratio of positron rates to the sum of the electron and positron
rates from DM annihilation in galactic halo, 
 is an  observable which allows discrimination among all the 
BSM  models we study: SUSY, LHT and UED.}
Even though these rates  will not be measured at the LHC,
we include these in our study, since they
will come from experiments with the same (LHC) time line, stressing
a very important complementarity between LHC and DM search experiments  
to decipher the underlying theory.
\end{itemize}

\noindent 
II) \underline{Heavy partner content}:
Even though LHT and UED are both "bosonic" supersymmetries, 
{their  heavy partner content differs significantly}. Since 
LHT has no heavy partner of gluon, one expects less QCD-induced events in the 
LHT scenario, as compared to SUSY and UED. 

\noindent
III) \underline {Existence of higher level modes}:
The higher level modes, e.g.,  the 2nd KK modes, appear only in UED 
scenarios and 
do not exist in SUSY or LHT models.
Hence, it is important to identify 
comprehensive particle spectrum as precisely as possible. 
These  measurements will be  affected by the experimental 
resolution of {\it all}  measurable quantities  viz. 
energy and momentum
of leptons, the jet energy as well as  the 
the missing transverse energy $\met$, hence  the calibration 
and alignment of the detectors.

\noindent
IV) \underline
{Majorana versus Dirac  nature of the heavy neutral
fermion partners}: The character of the heavy neutral fermions is 
clearly an important distinguishing feature among these models.
In LHT or UED models Majorana fermions are absent, whereas 
in all usual formulation of supersymmetric theories,
neutralinos and gluinos are Majorana fermions~\footnote{Note however, that
there exists a class of SUSY theories with Dirac gaugino 
masses ~\cite{Fox:2002bu,Kribs:2007ac} where this distinction between 
supersymmetric and nonsupersymmetric models may not hold true.}.
These  serve as a source of like-sign lepton signatures.
One of the observables which reflects this difference is the
$N_{l^+l^+} / N_{l^-l^-}$ ratio as well as the
ratio between multilepton rates and just $\met + jets$, viz., 
$R = \frac{N(\met + \mbox{jets})}{N(l's + \met + \ \mbox{jets})}$.
{In case of LHT and UED,  the  $N_{l^+l^+} / N_{l^-l^-}$ ratio 
is fixed  by parton density functions and the mass
of the heavy quarks produced in the t-channel reactions
initiated by  two valence  quarks in the initial state.
For example, this  ratio is between  3.5 and 5 for
the respective heavy quark mass between 0.3 and 1 TeV~\cite{Belyaev:2006jh},  
while in SUSY this ratio
is diluted by the same sign leptons originating from cascade gluino decays.
The ratio $R$ mentioned above is  larger in SUSY
compared to  LHT because of the presence of gluino in SUSY models. 
We plan to study this ratio for the case of UED
scenario.}
{The systematics of these measurements will be affected by the lepton charge
mis-identification probability and any lepton sign dependent systematics.} 

\noindent 
V) \underline
{b-jet and $\tau$ multiplicity}: 
For example, in the SUSY Focus Point region the b-jet multiplicity 
is enhanced due to Higgsino nature of neutralino
and suppressed mass of the lightest stop-quark
as compared to the first and second squark generations.
In fact top multiplicity may
also be used
effectively.
{This measurement will  be strongly affected
by  the b- and $\tau$-tagging efficiency and purity.}

\noindent
VI)\underline
{Single production of the heavy partner of the top}:
In LHT  single heavy top production is possible. 
Also  single KK2 (2nd KK mode) heavy top can be produced  
through KK2 parity violating coupling in UED (~\cite{Datta:2005zs}). 
 There is no such analog in SUSY.

\noindent
VII) \underline 
{The number of DM coannihilation channels.}
The number of DM coannihilation channels in the early Universe
can be considerably larger in the case of UED
scenario as compared to SUSY or LHT scenarios.
The set of UED coannihilating channels can include 
coannihilation  of KK photon with KK leptons, KK quarks, 
KK scalars, KK W/Z and KK gluons simultaneously.
This degeneracy then would lead to 
an enhanced number of decays of soft particles, resulting 
from several degenerate states.

\noindent
VIII)\underline 
{Various kinematical observables.}
We  will also include possible significant 
kinematical variables, some of which have been analyzed
in  previous studies.
\begin{itemize}
\item
number of leptons versus number of jets counts
including same-sign and opposite sign leptons
of various flavours
\item
invariant and transverse masses of multilepton
states
\item
kinematical edges
\item
event topology, including event shape variables as acoplanarity,
sphericity
\end{itemize}

{The comparison of generic features of  SUSY, LHT and 
UED stated above} is summarized in  Table~\ref{tab:comparison} 

\begin{table}[h]
\caption{Discriminating signatures between SUSY (MSSM), LHT and UED.
      See description in the text. "YES" or "NO" mean presence or absence of the
      particular signature respectively, "SS" stands for "like-sign leptons".} 
\label{tab:comparison}
\vspace*{0.2cm}  
\begin{tabular}{|l|l|l|l|l|}
\hline
\hline
\multicolumn{2}{|l|}{Variables} & SUSY (MSSM) & LHT  & UED \\ 
\hline
\hline 
\multicolumn{2}{|l|}{}      & heavy partners    &  heavy partners  	  &  heavy partners  	\\
\multicolumn{2}{|l|}{Spin}  & differ in spin	&  have the same   	  &  have the same   	\\
\multicolumn{2}{|l|}{}      & by 1/2	     	&  spin, no heavy      	  &  spin  	 	\\
\multicolumn{2}{|l|}{}      & 	     	     	&  gluon    	  	  &  	     	 	\\
\hline 
\multicolumn{2}{|l|}{Higher level}           	&NO   	     		& NO  	  	          &  YES 	\\
\multicolumn{2}{|l|}{modes}	  		& heavy partners     &  heavy partners  	  &  heavy partners  	\\
\hline 
\multicolumn{2}{|l|}{$N_{l^+l^+} / N_{l^-l^-}$} 	& $R_{SUSY} < R_{LHT}$ & $ R_{LHT}$ 	& $R_{UED}\simeq R_{LHT}$         		\\
\hline 
\multicolumn{2}{|l|}{}			& from several 		& only from 	&  only from 	      \\
\multicolumn{2}{|l|}{SS leptons rates}	& channels:     	& SS heavy	&  SS heavy  	      \\
\multicolumn{2}{|l|}{}			& SS heavy fermions,    & fermions 	&  fermions  	      \\
\multicolumn{2}{|l|}{}			&  Majorana fermions     & 		& 	   	\\
\hline 
\multicolumn{2}{|l|}{
$R = \frac{N(\met + jets)}{N(l's + \met + \ \mbox{jets})}$
} 
                & $R_{\rm SUSY}$ & $R_{\rm LHT} < R_{\rm SUSY}$ &  $R_{\rm UED}$		\\ 
\multicolumn{2}{|l|}{       }         	& 	& 	& to be studied  		\\ 
\hline 
\multicolumn{2}{|l|}{b-jet multiplicity} & enhanced (FP) & not enhanced		&   not enhanced 	\\ 
\hline     
\multicolumn{2}{|l|}{Single heavy top} 		& NO 			& YES		& YES  	\\
\multicolumn{2}{|l|}{	} 		     	& 			& 		& via KK2 decay 	\\
\hline
\multicolumn{1}{|l|}{polarization}       & $tt+\met$  &    to be studied           	&	to be studied	   &to be studied\\
\multicolumn{1}{|l|}{effects}		 & $\tau\tau+\met$  &    to be studied        &	to be studied	   &to be studied
	     	\\  
\hline 
\multicolumn{2}{|l|}{           }	&             &      		& typically low for\\ 
\multicolumn{2}{|l|}{Direct DM  }	&  high (FP)  & low  		& $\gamma_1$(5D) DM~\cite{Arrenberg:2008wy}\\ 
\multicolumn{2}{|l|}{detection rate} 	& low (coann) &(Bino-like LTP)	&typically high for \\
\multicolumn{2}{|l|}{}             	&    		& 	      	&$\gamma_H$(6D) DM~\cite{Arrenberg:2008wy}\\
\hline
\hline 
\end{tabular} 
\end{table} 

\section{Experimental issues}

Before one embarks on the study of distinguishing
among the BSM models, one will have to also establish how well these chosen
signals can be discriminated from the SM backgrounds. This will be 
an inherent part of our study. 
{The experimental issues involved in the signal extraction are related to 
the missing $E_T$ measurement, the reconstruction of hadronic, b and $\tau$ jets,
and the lepton identification, which are discussed here.}

Missing $E_T$ ({$\met$}) is primarily reconstructed from the energy deposits in the calorimeter and 
the reconstructed muon tracks. Apart from the hard scattering process of 
interest, many other sources, such as the underlying event, 
multiple interactions, pileup and electronic noise lead to energy
deposits and/or fake muon tracks. Classifying these energy deposits into various 
types (e.g. electrons, taus or jets) and calibrating them accordingly,
is the essential key for optimal { $\met$} measurement. In addition, the 
loss of energy in regions of inactive material and  dead detector
channels make the 
{ $\met$} measurement a real challenge.

The {$\met$} reconstruction algorithm starts from the energy deposits in 
calorimeter cells or clusters of cells (``raw { $\met$}''). 
The raw { $\met$} is then cleaned up from a
number of sources of fake { $\met$} : hot cells, overlay of beam-halo,
cosmics, detector malfunctions, detector hermiticity. Overall, the reconstruction
of { $\met$} is a challenging task and it  requires a good understanding of
the calorimeter response and  the { topology} of different signatures.
The { $\met$} resolution roughly scales with $\sqrt{\sum E_T}$, where
$\sum E_T$ is the scalar sum of the energies of the particles in the final
state, for  $\sum E_T<1.5$~TeV.

For the reconstruction of hadronic jets, a seeded fixed-cone reconstruction algorithm with a cone size
$\Delta R=\sqrt{\Delta\phi^2+\Delta\eta^2}=0.4$  is presently used for
search studies for  BSM physics. 
For future studies also the SISCone (Seedless Infrared Safe Cone) jet 
algorithm and the fast $K_{\rm T}$ algorithm are considered.
If one neglects the noise term, the jet energy resolution
varies between $50-100\%/\sqrt{E(\gev)}$.
Both experiments have strong
capabilities for the  identification of b-jets and $\tau$-jets in wide range of
transverse momentum for $\left|\eta\right|<2.5$. For a b-tagging
efficiency of $60\,\%$ and transverse momentum $20<p_T<100\,\gev$ a rejection of
above 100 and about 10 may be achieved against light and c-jets,
respectively, { with degradation of the performance for $p_T>100\,\gev$.}
For a  $\tau$-jet
efficiency of $50\,\%$, the rejection against hadronic jets  improves with $p_T$,
reaching rejection values of $O(10^2)-O(10^3) GeV$. 

Electrons are reconstructed as objects that have a track in the inner
tracker and an electromagnetic cluster in the EM calorimeter. 
The calorimeter is
designed to contain almost all of the energy of a high $p_T$ (TeV range) electron, and has
an energy resolution of $2-10\%/\sqrt{E(\gev)}$, depending on the experiment.
The inner tracker has an intrinsic $p_T$ resolution of a few times $10^{-4} p_T(TeV/c)$,
which is limited by early bremsstrahlung in its material. In order to separate isolated electrons 
originating from interesting events, from QCD background
(hadrons, jets and photons) with similar topology, several of their characteristics
are exploited. The EM cluster in the calorimeter is required to match with a track in the inner tracker
and the ratio of its energy over its momentum measured by the tracker (E/p) to be that of an electron. 
Cuts on the longitudinal (and lateral) shape of the shower
are applied, and minimal energy is allowed to be deposited in the Hadronic Calorimeter.


Muons are reconstructed as objects that have a track in the muon spectrometer and a
corresponding ("matched") track in the inner tracker. 
In the case of ATLAS, the good resolution of the muon spectrometer provides the possibility to
trigger and reconstruct muons in "stand alone" mode (no matching with the inner detector involved). 
The momentum resolution is maintained high for both
experiments. For muon $p_T$ in the TeV range the resolution 
 is limited by detector alignment in the
case of ATLAS and can be kept at $\sigma/p_T \approx 10\% $, whereas in the
case of CMS it is limited by energy losses in the iron yoke, and 
 it varies between $15-30\%$. In combination with the inner detector
track the resolution is improved to $5\%$. The muon detection and reconstruction efficiencies
for both experiments are high (above $95\%$). The charge misidentification probability
varies between $10^{-3} - 10^{-2}$ for muons below $100 \gev p_T$ and between $10^{-2}$ to
few times $10^{-1}$ for muons above $500\,\gev$, increasing with rapidity. Finally the
expected fake rate for muons, even for the high luminosity case, can be maintained to the
$\%$ level, while it is  an order of magnitude lower for low luminosity.

\section{Strategy}
{For the signature analysis
we will investigate details of each particular class of models as discussed 
above.
A set of significant signatures (the aim of our study)
for each model will be classified  
as shown in Table~\ref{tab:comparison}.
For example, for  MSSM a preliminary and still
incomplete version of such a classification is shown in
Table~\ref{tab:signatures}. 
\begin{table}[h]
\caption{DM motivated models and signatures. Only the MSSM is listed here.
         The following signatures: $\met$+jets, top polarization,
	 top-quark multiplicity are planned to be studied.
	 OSL and SSL stand for opposite-sign leptons
	 and same-sign leptons respectively.
	 }
\label{tab:signatures}
\vspace*{0.2cm}  
\begin{tabular}{|l|c|c|c|c|}
\hline
\hline
Signatures    and    Observables &\multicolumn{4}{c|}{SUSY (MSSM)} \\
\hline
& Focus point & Coann.\ & A-res. & Bulk 	\\ 
\hline 
\hline
1 $\ell$ + jets + $\met$ 	&  YES	&  YES	&  YES	&  YES	\\
\hline
OSL + jets + $\met$  	& YES & YES & YES & YES \\
\hline
SSL + jets + $\met$  	& YES & YES & YES & YES \\
\hline
3 $\ell$ + jets + $\met$  	& YES & YES & YES & YES \\
\hline	
4 $\ell$ + jets + $\met$  	& YES & YES & YES & YES \\
\hline
N b-jets 		& enhanced 
                              & YES & YES & YES \\
\hline 
H+$\met$+jets from cascades & & & & \\
\hspace*{1cm}$H \to \gamma\gamma, b\bar{b}$ 
			& YES & NO & NO & NO 	\\
\hspace*{1cm}$H \to VV, t\bar{t}$ 
                        & NO & NO & NO & NO 	\\
\hline
soft taus       	& YES & enhanced &  NO & NO \\
\hline
tau polarization  	& YES  &  YES  &   YES    &YES  \\
\hline
$N_{l^+l^+} / N_{l^-l^-}$   & $\sim 1:1$ & $<R_{LHT}$& $<R_{LHT}$ & $\ll R_{LHT}$ \\
\hline
DD rates, $\sigma (Z1 p)$  & enhanced & suppr. & suppr. 
& part. enhanced \\
\hline
ID rates, $\langle \sigma v\rangle (v\to 0)$   & enhanced & suppr. & suppr. 
& part. enhanced\\ 
\hline
\hline
\end{tabular}
\end{table}

Every "YES" entry in the Table 
means that the particular final state has the potential of
being able to discriminate among (or pinpoint to) different regions of the MSSM
space, consistent with DM constraints. 
For example, while the b-jets multiplicity (N b-jets in the Table)  
may allow to separate the SUSY signal
from the SM in all the regions of the MSSM parameter space,
 the amount of enhanced b-jet multiplicity is  very large
particularly  in the Focus Point region.

For mSUGRA, for example, the polarization of $\tau$ leptons produced
in the decay of 
$\tilde \tau_1\tilde\tau_1$ can be used very effectively to sharpen
up SUSY signature; for the coannihilation region where one expects soft
$\tau$'s, the fact that $\tau$'s from SUSY decays are polarized, can be used
very effectively to reduce SM background from the soft QCD jets.

Another example of the powerful discrimination between different DM motivated SUSY
regions are the
 dark  matter  direct detection  (DD)  rates which are proportional to 
neutralino scattering cross section off the nuclei, usually expressed
in terms of  $\sigma (Z1 p)$
as well as indirect dark matter detection rates (ID)
related to average of DM annihilation rate times velocity
in zero velocity limit,  $\langle \sigma v\rangle (v\to 0)$.

For each entry with a ``YES'' in the Table, the most important
contributing processes will be listed and studied in more detail. Similar
Tables will be  worked out for the LHT and UED.

In the very recent work~\cite{Hubisz:2008gg}
the authors aimed to distinguish  a quite specific
benchmark points for these theories with high cross section
in the first month of the LHC run.
We plan on
using analogous Tables for LHT and UED models
together with Table of "comparison", Table~\ref{tab:comparison}
to create a
``dictionary of LHC signatures'' and
examine a strategy to discriminate  all three classes of theories
{\it for  generic cases of their realization in wide range of the model space}.
This will be  the main difference and novelty of our study
in comparison with earlier ones.

\end{document}